\documentclass[10pt]{article} 

\usepackage[preprint]{tmlr}

\usepackage{amsmath,amsfonts,bm}

\def\eqref#1{equation~\ref{#1}}

\def\1{\bm{1}}

\DeclareMathAlphabet{\mathsfit}{\encodingdefault}{\sfdefault}{m}{sl}
\SetMathAlphabet{\mathsfit}{bold}{\encodingdefault}{\sfdefault}{bx}{n}

\usepackage{hyperref}
\usepackage{url}
\usepackage{booktabs}
\usepackage{graphicx}

\title{One Waveform, Two Stories: \\Benchmarking Speech–Acoustic Conflict in Audio-Language Models}

\author{\name Wenbin Wang \email wenbin.wang.cs@gmail.com \\
      \addr The University of Osaka
      \AND
      \name Yuheng Li \email li2464@wisc.edu \\
      \addr University of Wisconsin-Madison
      \AND
      \name Koichiro Yoshino \email yoshino.koichiro.es@osaka-u.ac.jp\\
      \addr The University of Osaka }

\def\month{MM}  
\def\year{YYYY} 
\def\openreview{\url{https://openreview.net/forum?id=XXXX}} 

\begin{document}

\maketitle
\raggedbottom

\begin{abstract}
Audio-language models (ALMs) are evaluated on their understanding of both speech and environmental sounds. What remains underexplored is how they respond when spoken claims contradict acoustic evidence within the same recording. We introduce SAC-Bench, a benchmark of intra-audio speech--acoustic conflict comprising 616,031 unique clips from 188 sound-event classes across seven semantic, temporal, and spatial tasks. It pairs environmental recordings with compatible or contradictory synthesized spoken claims and includes audio-only controls where applicable. Across five open-weight ALMs evaluated on semantic and temporal tasks, high audio-only accuracy often coexists with large losses under conflicting claims; Event Identity accuracy is 24.8--46.5 percentage points lower in conflict than compatible conditions. Temporal Order failures persist even when speech follows the complete event sequence, so simultaneous speech--event masking cannot fully explain them. A spatial case study with Spatial-Omni finds lower motion accuracy associated with both spoken claims and the speech source's physical trajectory. Prompt interventions sometimes restore acoustic accuracy, but can also reduce claim-aligned answers without improving grounding or worsen performance. SAC-Bench thus separates acoustic competence, conflict robustness, and intervention effects.
\end{abstract}

\begin{figure*}[!htbp]
    \centering
    \includegraphics[width=0.95\textwidth]{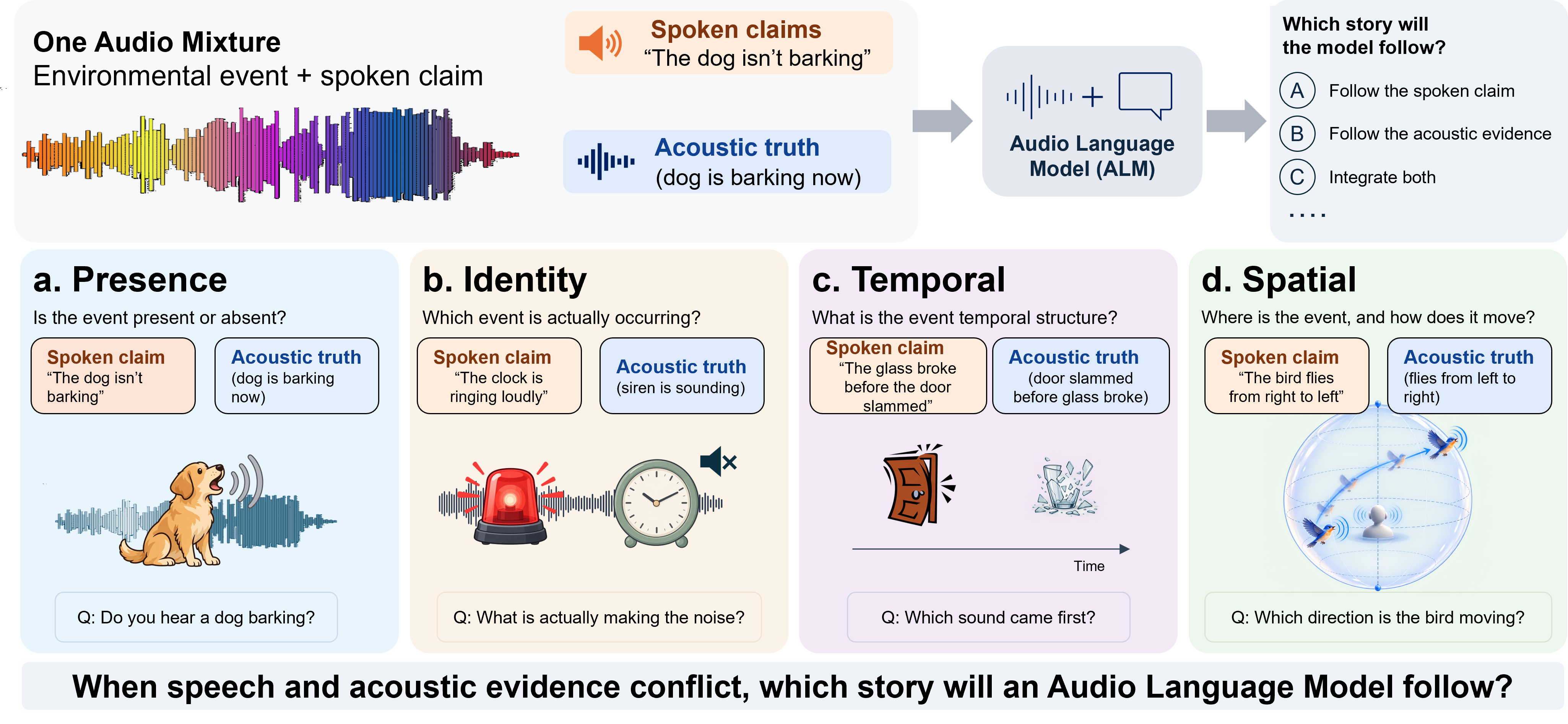}
    \caption{Overview of SAC-Bench. A single audio waveform contains spoken claims and environmental acoustic evidence that may imply conflicting interpretations. SAC-Bench evaluates this conflict across four task groups: event presence, event identity, temporal structure, and spatial configuration.}
    \label{fig:teaser}
\end{figure*}

\section{Introduction}

Audio-language models (ALMs) can process spoken language alongside non-speech acoustic information, including environmental sounds~\citep{tang2024salmonn,xu2025qwen3,ghosh2026audio}. Unlike cascaded dialogue systems, whose downstream components largely rely on ASR transcripts~\citep{siri2017heysiri}, ALMs can in principle use acoustic evidence that spoken words alone do not convey. However, making this evidence available to a model does not establish whether it actually contributes to the model's predictions.

This distinction is difficult to assess when spoken claims agree with the environmental acoustic evidence. If a speaker says that a dog is barking and a bark is audible, a model can answer correctly by following the spoken claim without recognizing the bark itself. The same ambiguity arises when speech accurately describes an event's timing or a sound source's location. Consequently, success under aligned conditions cannot distinguish faithful acoustic grounding from reliance on linguistic shortcuts~\citep{xiong2026deaf}. Figure~\ref{fig:teaser} illustrates how conflicting spoken claims and acoustic evidence are instantiated across the four task groups.

Conflict-based evaluation provides a direct way to expose such shortcut behavior. Prior work has studied contradictions between visual and textual evidence, audio and video, and audio and externally supplied text, revealing systematic preferences among competing information sources~\citep{zhang2025robust,sung2024avhbench,zhong2026have,cho2026wins,billa2026audio}. These settings primarily concern inter-modal conflict, where contradictory evidence is provided through separately represented modalities. Speech-centered audio creates a distinct setting: the same audio input may contain foreground linguistic claims together with non-linguistic evidence about the surrounding events, their temporal structure, and their spatial configuration. We refer to this setting as \textbf{intra-audio speech--acoustic conflict}, where spoken claims and acoustic evidence within the same audio input imply incompatible interpretations of what occurred, when it occurred, or where and how a sound source is situated or moves.

Recent benchmarks have separately investigated several components of this problem. Intra-audio conflict benchmarks primarily study lexical–prosodic inconsistencies, speaker attributes, or semantic background events~\citep{cheng2026aha, huang2026tone, xiong2026deaf}. Temporal audio benchmarks evaluate temporal ordering, duration, or timestamp grounding~\citep{ghosh2024compa,bhattacharya2025benchmarking,sun2026spotsound}, while spatial-audio benchmarks test source localization, distance, spatial relations, and motion trajectories~\citep{zheng2024bat,jiang2026sci,zhu2026spatial,sun2025audiomotionbench}. Yet existing intra-audio conflict evaluations primarily focus on semantic or paralinguistic inconsistencies, whereas temporal and spatial benchmarks generally assess acoustic capabilities without contradictory spoken claims. These settings leave underexplored how spoken claims influence grounding across event semantics, temporal structure, and spatial configuration. Spatial grounding additionally requires distinguishing the environmental target's location or trajectory from both the spoken claim and the speech source's own spatial cues.

We introduce the Speech--Acoustic Conflict Benchmark (SAC-Bench), with 616,031 unique audio clips across seven tasks spanning semantic, temporal, and spatial grounding. It combines compatible and conflicting spoken claims with single-modality controls where applicable, while matched spatial conditions vary claim content and speech-source geometry. Across five ALMs on semantic and temporal tasks, strong audio-only recognition coexists with substantial speech-aligned errors. Temporal Order failures persist even when speech follows the complete event sequence. A spatial case study further reveals sensitivity to both semantic and physical speech cues. Prompt interventions show that reducing speech following does not necessarily improve acoustic accuracy.

\section{Related Work}

\subsection{Multimodal and Intra-Audio Conflict}

Conflicting vision--text and audio--visual inputs reveal systematic preferences among competing evidence sources~\citep{zhang2025robust,sung2024avhbench,chen2026avid}. In audio--text settings, MCR-BENCH and ALME document reliance on externally supplied text despite contradictory audio~\citep{wang2025audio,billa2026audio}. Mechanistic analysis further shows that textual pathways can suppress intact audio representations during late-layer integration~\citep{cho2026wins}. These findings motivate distinguishing available acoustic information from its use in prediction, although mechanisms identified under inter-modal conflict need not transfer directly to intra-audio settings.

Within a single waveform, DEAF evaluates conflicts involving emotional prosody, background scenes, and speaker attributes, separately varying spoken-content conflict and misleading textual prompts~\citep{xiong2026deaf}. Its background-sound task asks which environment the sounds suggest, providing a direct precedent for environmental speech--acoustic conflict at the scene level. VoxParadox instead targets paralinguistic attributes through contradictions between transcript claims and vocal delivery~\citep{pang2026audio}. Audio Hallucination Attacks introduces AHA-Eval, combining misleading queries with injected speech describing absent events~\citep{seth2026audio}. These studies establish overlapping forms of linguistic interference and motivate examining its variation across more structured environmental judgments.

\subsection{Audio Grounding Benchmarks}

General suites such as AIR-Bench, AudioBench, MMAU, and MMAR assess broad audio understanding and reasoning~\citep{yang2024air,wang2025audiobench,sakshi2025mmau,ma2026mmar}. Reliability-oriented evaluations examine unsupported responses: AHa-Bench covers semantic, acoustic, and semantic--acoustic confusion, including speaker-attribute contradictions and speech-induced inference of absent sounds~\citep{cheng2026aha}, while HalluAudio evaluates hallucinations across speech, sounds, and music~\citep{zhao2026halluaudio}. Such coverage complements conflict-specific evaluation; task accuracy and hallucination rates alone do not characterize robustness to systematically varied spoken claims.

Temporal benchmarks probe event structure: CompA-order tests occurrence and ordering, TREA evaluates ordering, duration, and counting, and SpotSound-Bench targets fine-grained event localization~\citep{ghosh2024compa,bhattacharya2025benchmarking,sun2026spotsound}. SpatialSoundQA/BAT evaluates binaural spatial perception, Sci-Phi describes spatial acoustic scenes, and SO-Bench and AudioMotionBench assess spatial relations or source motion~\citep{zheng2024bat,jiang2026sci,zhu2026spatial,sun2025audiomotionbench}. These tasks characterize temporal and spatial capabilities without systematically crossing them with contradictory spoken descriptions.

SAC-Bench builds on these conflict and capability evaluations through seven environmental-event tasks. Available audio-only controls measure performance without speech; non-overlapping Temporal Order sequences test conflict beyond simultaneous masking. Spatial conditions separately specify the environmental target, spoken claim, and physical speech source, enabling matched comparisons of semantic and physical interference. Table~\ref{tab:benchmark_comparison} distinguishes event-level conflict coverage from scene-level and paralinguistic conflict; its coverage markers describe evaluation designs, not model performance.

\begin{table*}[htbp]
\centering
\caption{Comparison of audio-language benchmarks in speech--acoustic conflict coverage and dataset scale.}
\label{tab:benchmark_comparison}
\vspace{3pt}
\setlength{\tabcolsep}{3.0pt}

\resizebox{\textwidth}{!}{%
\begin{tabular}{@{}ll*{7}{c}ccc*{4}{r}@{}}
\toprule
& &
\multicolumn{7}{c}{\textbf{Speech--acoustic event conflict}}
& & & &
\multicolumn{2}{c}{\textbf{Audio}} &
\multicolumn{2}{c}{\textbf{QA pairs}} \\
\cmidrule(lr){3-9}
\cmidrule(lr){13-14}
\cmidrule(l){15-16}

\textbf{Benchmark} &
\textbf{Grounding tasks} &
\textbf{P} &
\textbf{I} &
\textbf{St} &
\textbf{Or} &
\textbf{Lc} &
\textbf{Sm} &
\textbf{Rm} &
\shortstack{\textbf{Other}\\\textbf{conf.}} &
\shortstack{\textbf{SF}\\\textbf{ctrl.}} &
\shortstack{\textbf{3-src}\\\textbf{GT}} &
\textbf{Total} &
\textbf{Eval} &
\textbf{Total} &
\textbf{Eval} \\
\midrule

CompA~\citep{ghosh2024compa}
& Order, attributes (600 inst.)
& -- & -- & -- & -- & -- & -- & --
& -- & -- & --
& 1,300 & 1,300 & n/a & n/a \\

AHa-Bench$^{\dagger}$~\citep{cheng2026aha}
& Presence, duration, order, distance
& -- & -- & -- & -- & -- & -- & --
& $\circ$ & -- & --
& 396 & 396 & 906 & 906 \\

HalluAudio~\citep{zhao2026halluaudio}
& Presence, overlap, order
& -- & -- & -- & -- & -- & -- & --
& -- & -- & --
& n/r & n/r & 5,720 & 5,720 \\

SO-Bench~\citep{zhu2026spatial}
& Onset, location, distance, motion
& -- & -- & -- & -- & -- & -- & --
& -- & -- & --
& 7,877 & 7,877 & 7,877 & 7,877 \\

AHA-Eval~\citep{seth2026audio}
& Absent-event queries
& $\bullet$ & -- & -- & -- & -- & -- & --
& -- & $\circ$ & --
& n/r & n/r & $\sim$6.5K & $\sim$6.5K \\

DEAF$^{\dagger}$~\citep{xiong2026deaf}
& Background, prosody, speaker
& -- & -- & -- & -- & -- & -- & --
& $\bullet$ & -- & --
& 2,756 & 2,756 & n/r & n/r \\

VoxParadox~\citep{pang2026audio}
& Paralinguistics, speaker
& -- & -- & -- & -- & -- & -- & --
& $\bullet$ & -- & --
& 2,000 & 2,000 & 2,000 & 2,000 \\

\midrule

\textbf{SAC-Bench (Ours)}
& \textbf{Seven event tasks}
& $\bullet$ & $\bullet$ & $\bullet$
& $\bullet$ & $\bullet$ & $\bullet$ & $\bullet$
& -- & $\bullet$ & $\bullet$
& \textbf{616,031}
& \textbf{223,021}
& \textbf{2,043,836}
& \textbf{744,086} \\

\bottomrule
\end{tabular}%
}

\vspace{3pt}
\begin{minipage}{\textwidth}
\scriptsize
\emph{Legend:}
$\bullet$ = evaluated; $\circ$ = partial; -- = absent; n/r = not reported; n/a = not applicable.

\emph{Conflict dimensions:}
P = presence; I = identity; St = state; Or = order; Lc = location; Sm = spherical motion; Rm = radial motion. Other = non-event conflict; SF = speech-free control; 3-src = independent acoustic, claim, and speaker geometry. Audio counts denote clips or stimuli; SAC-Bench counts each distinct final audio path once across task references; QA counts include diagnostics but exclude prompt variants. CompA reports matching instances; DEAF reports stimuli.

\smallskip
$^{\dagger}$DEAF's BSC task asks which environment the background sounds suggest. Its scene-level speech--acoustic conflict is marked separately from event-label Presence and Identity. AHa-Bench has partial Presence coverage through speech-induced inference of absent sounds.
\end{minipage}

\end{table*}

\section{Dataset Construction Method}
\label{sec:dataset_construction}

\begin{figure*}[t]
    \centering
    \includegraphics[width=0.95\textwidth]{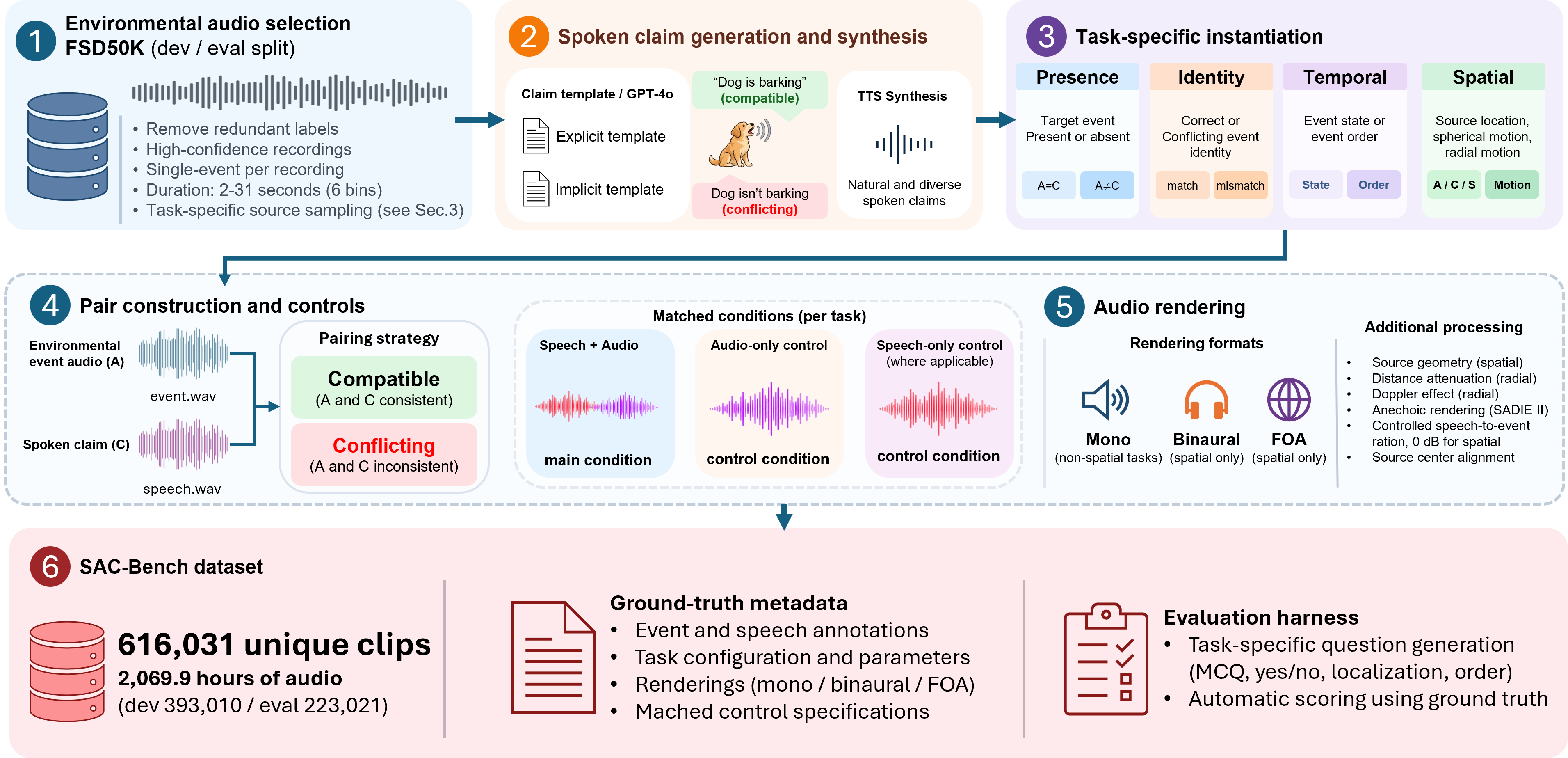}
    \caption{SAC-Bench construction pipeline, from environmental audio selection and spoken claim synthesis to conflict generation, audio rendering, and evaluation.}
    \label{fig:construction_overview}
\end{figure*}

Figure~\ref{fig:construction_overview} summarizes our data construction pipeline. Each task pairs environmental acoustic evidence ($A$) with a spoken claim ($C$) that may agree or conflict with it. The evaluated answer concerns the environmental event. For example, a dog-barking recording paired with a spoken claim about barking forms a compatible condition, whereas the same recording paired with a claim about a siren forms a conflicting condition. Temporal State also uses the speech segment as a temporal reference; spatial tasks additionally distinguish the claim $C$ from the speech source's physical geometry ($S$). In figures and relation labels, $=$ and $\neq$ denote matches or mismatches in task-relevant attributes, not identity among audio signals, spoken claims, and speech sources.

We curate FSD50K recordings~\citep{fonseca2021fsd50k} by removing redundant ancestor labels and retaining recordings with one remaining event label and unanimously positive available PP/PNP ratings. We preserve the original splits and evaluate models only on the exhaustively annotated evaluation split, treating unannotated in-vocabulary events as absent under its annotation protocol. Recordings span six duration bins from 2 to 31 seconds, with at least five clips per class within each split and bin. Claims are generated with GPT-4o or task-specific templates and synthesized with F5-TTS~\citep{hurst2024gpt,chen2025f5}. Synthesizing speech lets us pair the same environmental recording with different claims while controlling speaker identity, timing, level, and spatial geometry across matched conditions. Recording natural speech for all such combinations would be harder to scale and standardize. Selection and synthesis details appear in Appendix~\ref{app:dataset_details}.

\subsection{Semantic Grounding: Event Presence and Identity}
\label{sec:semantic_grounding}

\textbf{Presence} asks whether a target event is audible; \textbf{Identity} asks which environmental event is audible. Claims vary in polarity and, where available, explicitness. Explicit claims name the target, implicit claims avoid its label and screened synonyms, and negative Identity claims deny the target without implying silence. Each claim is paired with the target event, no environmental event, or a near/far alternative. CLAP text embeddings~\citep{wu2023large} define a dissimilar candidate pool, within which near and far alternatives are selected using Lin similarity on the AudioSet ontology graph. Positive claims agree only with the target event; negative claims conflict only when it is present. Speech is center-aligned with the event at speech-to-event RMS ratios of $\{-10,-5,0,5,10\}$ dB. No-event conditions contain speech only; audio-only controls preserve the environmental recording. See Appendices~\ref{app:event_selection} and~\ref{app:semantic_rendering}.

\subsection{Temporal Grounding: Event Timing and Order}
\label{sec:temporal_grounding}

\textbf{Temporal State} places an event before or after a spoken timing claim. Speech provides the temporal reference, so this task has no audio-only control. \textbf{Temporal Order} presents two source recordings in their claimed or reversed order, with audio-only controls for both sequences. Speech follows the complete event sequence, avoiding simultaneous speech--event masking; inter-event and speech--event gaps are 0.2 seconds. See Appendices~\ref{app:event_selection} and~\ref{app:temporal_rendering}.

\subsection{Spatial Grounding: Source Location and Motion}
\label{sec:spatial_grounding}

\textbf{Static Location}, \textbf{Spherical Motion}, and \textbf{Radial Motion} use 38 spherical Voronoi zones for direction~\citep{renka1997algorithm}. Five distinct recordings per eligible class and split are reused across matched conditions. The speech-source geometry $S$ is undefined for non-spatial speech.

\textbf{Static Location} samples claims from five angular-distance strata relative to $A$: Same, Near, Medium, Far, and Antipodal. For this task, $A$ is the acoustic ground-truth zone, $C$ is the zone specified by the spoken claim, and $S$ is the zone assigned to the physical speech source. Spatialized speech uses five relations among these zones: L1 (all three match), L2 (only $A$ and $C$ match), L3 (only $A$ and $S$ match), L4 (only $C$ and $S$ match), and L5 (all three differ). One L1, one L2, four of each L3--L5, and five non-spatial conditions yield 19 conditions per source.

\textbf{Spherical Motion} follows a constant-radius great-circle arc. Claims describe its actual, reversed, or midpoint-orthogonal trajectory; speech is non-spatial or follows one of these three trajectories, yielding $3\times4=12$ conditions. \textbf{Radial Motion} crosses two acoustic directions (approaching or receding), two claim directions, and three speech geometries (non-spatial, same direction, or opposite direction), yielding $2\times2\times3=12$ conditions. Both motion tasks require recordings of at least 4 seconds.

Environmental events and spatialized speech are rendered in binaural and FOA formats with anechoic propagation, no added synthetic noise, and 0-dB active-RMS SER. Audio-only controls preserve the rendered environmental event. See Appendices~\ref{app:conflict_details} and~\ref{app:spatial_rendering} for sampling and rendering details.

\subsection{Dataset Statistics}

\begin{figure*}[t]
    \centering
    \includegraphics[width=0.85\textwidth]
    {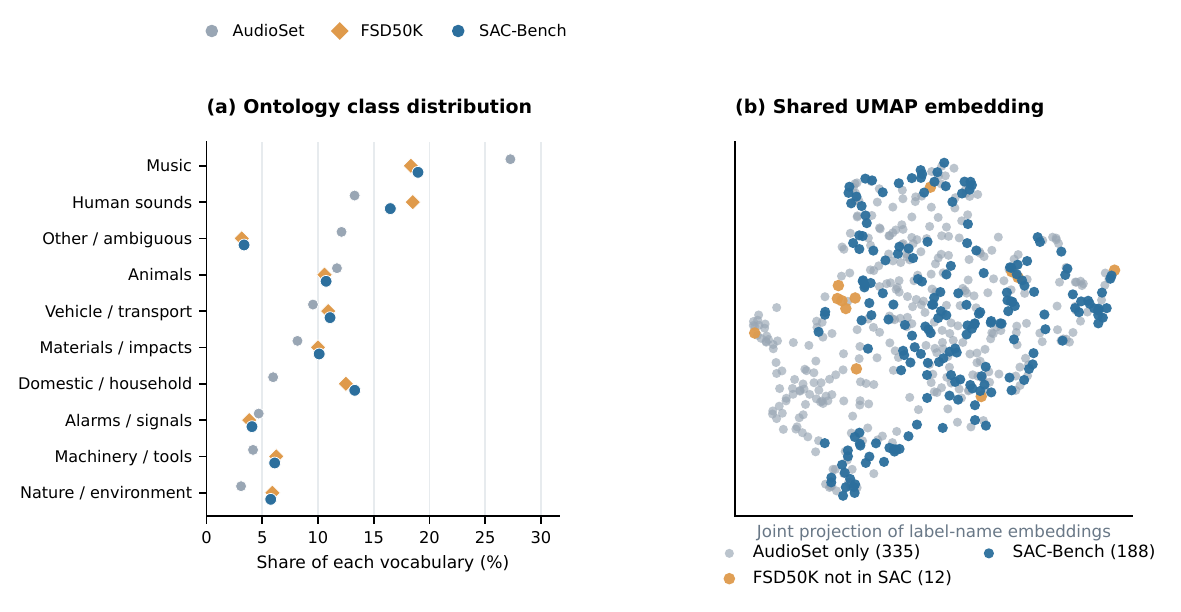}
    \caption{Taxonomic and semantic coverage of sound-event labels in AudioSet, FSD50K, and SAC-Bench. (a) Label distribution across ten ontology categories with fractional assignments. (b) UMAP projection of label-name embeddings.}
    \label{fig:semantic_coverage}
\end{figure*}

SAC-Bench contains 616,031 unique audio clips (2,069.9 hours), derived from 16,340 environmental recordings spanning 188 sound-event classes (94\% of FSD50K). The development and evaluation splits contain 393,010 and 223,021 clips, respectively. Table~\ref{tab:dataset_statistics} reports task-level counts and their global union; Figure~\ref{fig:semantic_coverage} characterizes event-vocabulary coverage. Each final audio path is counted once. Source recordings, synthesized speech identifiers, and final audio paths are disjoint across splits. Modality breakdowns, source durations, text overlap, and counting details appear in Appendix~\ref{app:dataset_statistics}.

\begin{table}[t]
\centering
\small
\caption{Unique audio clips by task and FSD50K split. The final row reports the deduplicated total across all tasks.}
\label{tab:dataset_statistics}
\vspace{5pt}
\begin{tabular}{lrrr}
\toprule
\textbf{Task} & \textbf{Dev} & \textbf{Eval} & \textbf{Total} \\
\midrule
Presence         & 160,663 & 81,892 & 242,555 \\
Identity         & 160,658 & 81,892 & 242,550 \\
Temporal State   &   2,884 &  1,520 &   4,404 \\
Temporal Order   &   4,304 &  2,258 &   6,562 \\
Static Location  &  36,346 & 28,200 &  64,546 \\
Spherical Motion &  17,620 & 14,638 &  32,258 \\
Radial Motion    &  18,930 & 15,764 &  34,694 \\
\midrule
\textbf{All tasks (union)}
& \textbf{393,010}
& \textbf{223,021}
& \textbf{616,031} \\
\bottomrule
\end{tabular}
\end{table}

\section{Experimental Setup}

Our evaluation tests whether ALMs use environmental evidence when spoken claims conflict with it. Where available, audio-only controls assess acoustic competence; compatible--conflict gaps describe performance across conditions; and speech-following rate (SFR) identifies claim-aligned responses. High audio-only accuracy with low conflict accuracy and high SFR indicates competence in isolation but frequent claim-aligned errors in mixtures. Prompt interventions are judged by both accuracy and SFR, because reduced speech following alone need not improve grounding. Matched spatial conditions distinguish claim content from speech-source geometry.

\subsection{Evaluated Models and Input Representations}
\label{sec:evaluation_models}

We evaluate five open-weight audio-language models: Qwen3-Omni-30B-A3B-Instruct, Step-Audio-2-mini, Audio Flamingo Next, MOSS-Audio-8B-Instruct, and Spatial-Omni SO30B~\citep{xu2025qwen3,wu2025step,ghosh2026audio,yang2026moss,zhu2026spatial}. All five are evaluated on presence, identity and temporal tasks; Spatial-Omni uses its mono pathway for these tasks. Spatial tasks are evaluated only with Spatial-Omni's four-channel first-order Ambisonics (FOA) encoder. FOA inputs longer than 20 seconds and binaural inputs are skipped. Checkpoints, preprocessing, and inference settings appear in Appendix~\ref{app:inference_configuration}.

\subsection{Evaluation Questions and Answer Choices}
\label{sec:evaluation_condition}

Presence poses Yes/No questions about the acoustic event and four alternative labels, adding the spoken target when needed and merging duplicates. Identity uses five choices containing the acoustic answer and, for positive conflicts, the speech-asserted identity; absent-event recordings include ``No environmental sound is present'' as the correct choice. Temporal tasks use binary choices. Spatial primary questions use 38 zones for Static Location, four directed trajectories for Spherical Motion, and two directions for Radial Motion.

Choices and their ordering are fixed within each matched source family across speech conditions, representations, and prompt policies. Separate spatial diagnostics assess the spoken claim, speaker geometry, their relation to the environmental source, and motion endpoints; these diagnostics are excluded from primary acoustic accuracy. Semantic analyses additionally stratify by claim formulation, polarity, and duration. Question and distractor construction appear in Appendix~\ref{app:question_construction}.

\subsection{Prompt Policies and Inference}
\label{sec:prompt_protocol}

We compare neutral prompting (no additional instruction), acoustic-grounded prompting (answer from environmental sounds), and conflict-aware prompting (also verify potentially false spoken claims). Temporal State instructions retain the speech segment as the temporal reference without treating its timing claim as evidence. Audio-only controls and diagnostic questions use neutral prompting.

Each zero-shot request contains the recording and a textual question, without demonstrations, transcripts, or reference answers. We use model-native input formats, disable sampling, and generate one response per question without retries or answer selection. Full instructions and inference settings appear in the appendix.

\subsection{Evaluation Metrics}
\label{sec:eval_matrics}

Completed responses form the scoring population, where unparseable outputs count as incorrect. Inference failures (e.g., skips, missing predictions) are excluded from the denominator.

\paragraph{Non-spatial performance.}
Identity and temporal tasks use question-level acoustic accuracy. Presence uses recording-level balanced accuracy: Yes and No recalls receive equal weight within each recording, and recordings receive equal weight. Conditions with only one reference class use recording-averaged recall for that class. Appendix~\ref{app:presence_metrics} gives the formal definitions.

\paragraph{Speech-following rate (SFR).}
For each conflicting speech--audio question, let $\mathcal S_i$ be the answer option or options implied by the spoken claim, derived directly from the claim and question. We include a completed question in $\mathcal D$ only when this speech-implied reference is defined and disjoint from the acoustically correct answer set. The speech-implied reference is derived from the claim annotation, not from model predictions on speech-only inputs. For example, negative Identity claims without a unique implied answer are excluded. SFR is the proportion of eligible mixed-input responses that select a speech-implied option:
\begin{equation}
\mathrm{SFR}
=
\frac{1}{|\mathcal D|}
\sum_{i\in\mathcal D}
\mathbb I[\hat y_i\in\mathcal S_i].
\end{equation}
A higher SFR indicates more claim-aligned responses under conflict, rather than merely more errors. A lower SFR does not by itself imply better acoustic grounding: the model may select another incorrect option or produce an invalid response. We therefore interpret SFR together with acoustic accuracy. 

\paragraph{Spatial performance and diagnostics.}
Spatial primary accuracy averages completed questions within each source recording, then recordings within each event class, and finally classes, preventing sources with more rendered variants from dominating. Random chance baselines are $1/38$, $1/4$, and $1/2$ for Location, Spherical Motion, and Radial Motion. Diagnostic questions targeting spoken claims or speaker geometry are scored separately. Conflict-target attraction measures the change in selecting a conflicting claim or speaker state relative to a matched audio-only control. Formal definitions appear in Appendix~\ref{app:chance_metrics}.

\paragraph{Comparisons and uncertainty.}
Paired comparisons use completed matched observations with identical answer choices and ordering. We report 95\% confidence intervals using source-cluster bootstrap for non-spatial tasks (connected components of shared sources for Temporal Order) and event-class bootstrap for spatial tasks, retaining recordings and their rendered variants within each resampled class.

\section{Results and Analysis}
\label{sec:results}

\subsection{Acoustic Competence and Speech-Aligned Failures}
\label{sec:overall_performance}

\begin{table}[t]
\centering
\small
\caption{Neutral-prompt acoustic performance and SFR (\%). Presence uses recording-level BAcc; other tasks use accuracy. Gap = compatible minus conflict (pp). SFR uses only eligible conflict questions.}
\label{tab:4_1_overall}
\vspace{5pt}
\begin{tabular}{lrrrrr}
\toprule
Model & Audio-only & Compatible & Conflict & Gap & SFR \\
\midrule
\multicolumn{6}{l}{\textit{Event Presence (BAcc)}} \\
MOSS-Audio-8B & 81.2 & 78.5 & 65.7 & 12.9 & 67.5 \\
Audio Flamingo Next & 84.0 & 85.2 & 71.1 & 14.1 & 67.8 \\
Qwen3-Omni & 91.3 & 89.0 & 79.3 & 9.7 & 64.2 \\
Step-Audio-2-mini & 85.8 & 79.9 & 67.8 & 12.0 & 74.7 \\
Spatial-Omni (SO-30B) & 73.1 & 72.8 & 69.3 & 3.5 & 64.6 \\
\midrule
\multicolumn{6}{l}{\textit{Event Identity (accuracy)}} \\
MOSS-Audio-8B & 46.7 & 57.4 & 29.1 & 28.4 & 60.9 \\
Audio Flamingo Next & 81.4 & 76.2 & 35.6 & 40.6 & 71.5 \\
Qwen3-Omni & 84.6 & 83.0 & 39.8 & 43.2 & 70.5 \\
Step-Audio-2-mini & 88.0 & 79.4 & 32.9 & 46.5 & 80.8 \\
Spatial-Omni (SO-30B) & 78.2 & 76.1 & 51.3 & 24.8 & 47.0 \\
\midrule
\multicolumn{6}{l}{\textit{Temporal Order (accuracy)}} \\
MOSS-Audio-8B & 94.0 & 94.2 & 43.6 & 50.7 & 56.4 \\
Audio Flamingo Next & 87.4 & 89.6 & 27.6 & 62.0 & 71.8 \\
Qwen3-Omni & 98.2 & 99.7 & 58.4 & 41.3 & 41.6 \\
Step-Audio-2-mini & 72.6 & 96.8 & 7.5 & 89.3 & 92.4 \\
Spatial-Omni (SO-30B) & 92.7 & 86.2 & 70.5 & 15.7 & 29.5 \\
\bottomrule
\end{tabular}
\end{table}

\paragraph{Acoustic competence without speech.} 
Table~\ref{tab:4_1_overall} compares three representative grounding tasks under neutral prompting. Conflict accuracy covers all scored conflict questions, whereas SFR requires a defined speech-implied answer. The columns therefore have different denominators and need not sum to 100\%, particularly for Identity. Without spoken claims, Qwen3-Omni achieves 84.6\% accuracy on Event Identity and 98.2\% on Temporal Order, while Step-Audio-2-mini reaches 88.0\% on Identity. These audio-only results establish substantial acoustic competence in several models. Compatible-condition performance alone cannot establish this competence, since the correct spoken claim may already provide the answer.

\paragraph{Performance under conflicting evidence.}
All five models score lower under conflicting than compatible speech, with gaps of 3.5--14.1 percentage points (pp) for Presence, 24.8--46.5 pp for Identity, and 15.7--89.3 pp for Temporal Order. These are descriptive condition-level gaps, not paired effects holding the environmental waveform fixed. Step-Audio-2-mini illustrates the pattern: its Temporal Order accuracy is 72.6\% with audio only and 96.8\% with compatible speech, but 7.5\% under conflict. Strong audio-only recognition thus does not guarantee high accuracy when speech contradicts the acoustic evidence.

\paragraph{Claim-aligned responses under conflict.}
Among conflicting five-choice Identity questions with a defined speech-implied answer,SFR ranges from 47.0\% to 80.8\%, above the 20\% uniform-choice reference. It reaches 80.8\% for Step-Audio-2-mini, 71.5\% for Audio Flamingo Next, and 70.5\% for Qwen3-Omni. Eligible conflict responses therefore frequently select the misleading spoken option. SFR describes this directional response pattern; it does not measure the fraction of all conflict errors caused by speech.

\subsection{Variation in Conflict Sensitivity}
\label{sec:conflict_heterogeneity}

Conflict sensitivity varies across speech level, claim formulation, and temporal task structure. We examine these factors before separately analyzing semantic and physical spatial cues.

\begin{figure*}[t]
\centering
\includegraphics[width=\textwidth]{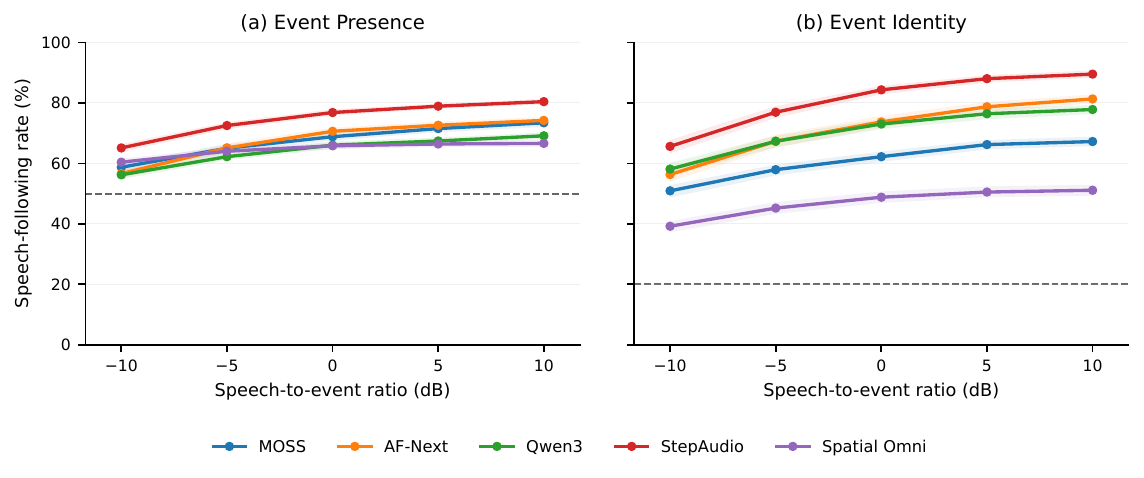}
\caption{SFR versus speech-to-event ratio (SER) under neutral prompting. Shading indicates source-cluster bootstrap 95\% confidence intervals; dashed lines indicate uniform-choice references.}
\label{fig:4_2_ser}
\end{figure*}

\begin{figure*}[t]
\centering
\includegraphics[width=\textwidth]{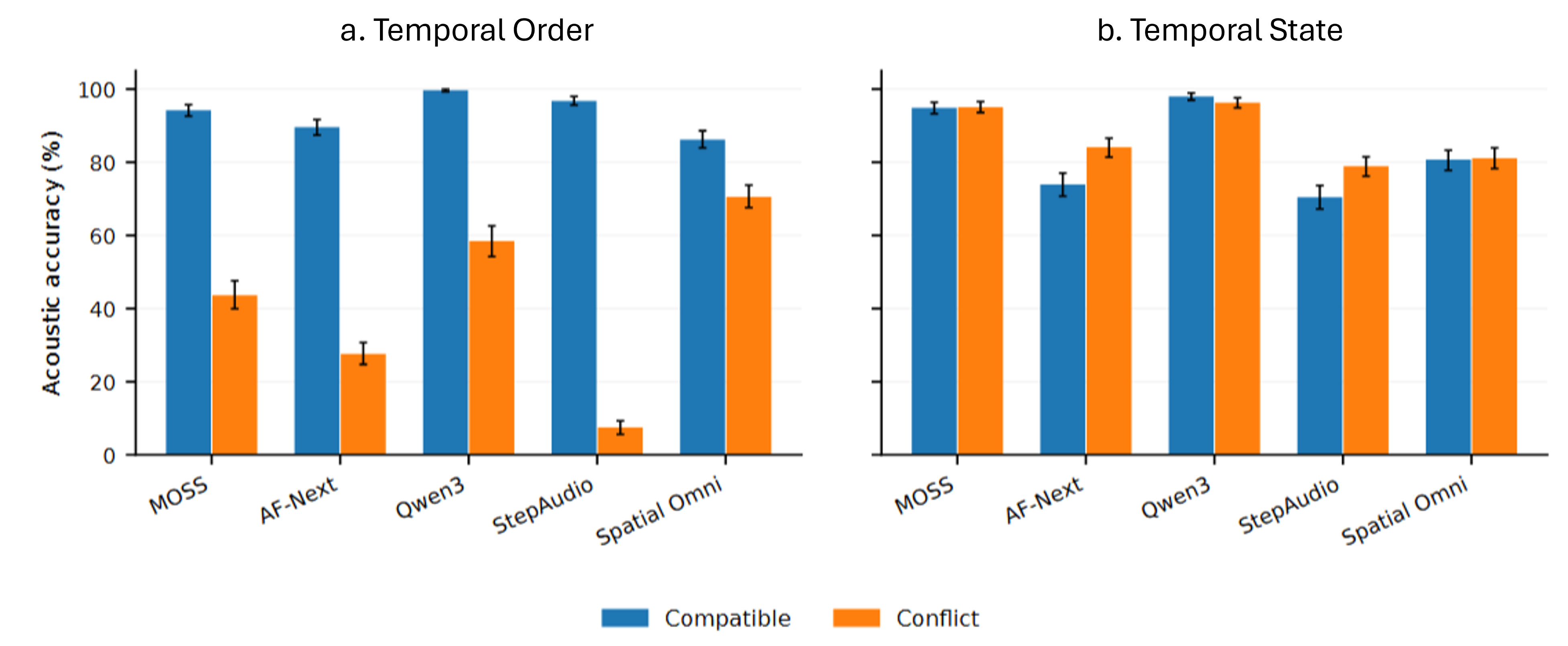}
\caption{Acoustic accuracy under compatible and conflicting claims for (a) Temporal Order and (b) Temporal State. Error bars show 95\% source-cluster bootstrap confidence intervals ($n=760$ per model--condition cell).}
\label{fig:4_2_temporal}
\end{figure*}

\paragraph{Relative acoustic prominence.}
Increasing the speech-to-event ratio (SER) from $-10$ to $+10$ dB increases SFR by 11.9--25.0 pp for Identity and 6.3--17.7 pp for Presence across the five models (Figure~\ref{fig:4_2_ser}). Even at $-10$ dB, Identity SFR remains 39.2--65.6\%, above the 20\% uniform-choice reference. Speech-aligned responses therefore persist when speech is quieter than the environmental event. However, SER changes both speech intelligibility and acoustic masking, preventing attribution of this trend solely to linguistic preference.

\paragraph{Claim formulation and direction.} 

Among positive conflicting Identity claims, explicit statements yield SFRs 13.7--29.1 pp higher than implicit statements over shared duration bins, including 79.8\% versus 50.7\% for Audio Flamingo Next. Presence also exhibits strong polarity asymmetries: positive versus negative claim SFR is 95.0\% versus 22.9\% for Qwen3-Omni and 95.4\% versus 1.2\% for Spatial-Omni. Because Presence claim polarity is coupled with target-event presence, these differences characterize condition-dependent behavior rather than an isolated wording effect.

\paragraph{Temporal task structure.} 

Temporal Order exhibits substantial conflict-associated accuracy decreases across all five models, whereas Temporal State shows no comparable pattern (Figure~\ref{fig:4_2_temporal}). In fact, compatible-to-conflict accuracy rises from 73.9\% to 84.1\% for Audio Flamingo Next and from 70.4\% to 78.9\% for Step-Audio-2-mini on Temporal State. Importantly, Temporal Order places the misleading speech after the entire environmental event sequence. Its failures therefore cannot be explained solely by simultaneous speech--event masking. Conflict sensitivity differs considerably even between tasks involving temporal acoustic information.

\subsection{Semantic and Physical Cues in Spatial Grounding}
\label{sec:spatial_results}

\begin{table}[t]
\centering
\small
\caption{Spatial-Omni acoustic accuracy (\%) on eligible FOA inputs under neutral prompting. NS/S denote non-spatial/spatialized speech; both columns pool compatible and conflicting conditions.}
\label{tab:4_2_spatial}
\begin{tabular}{lrrrr}
\toprule
Task & Chance & Audio-only & Speech (NS) & Speech (S) \\
\midrule
Location & 2.6 & 31.3 & 21.0 & 21.5 \\
Motion   & 25.0 & 50.5 & 35.0 & 33.2 \\
Radial   & 50.0 & 50.3 & 49.2 & 49.7 \\
\bottomrule
\end{tabular}
\end{table}

\paragraph{Task-dependent acoustic competence.}
Table~\ref{tab:4_2_spatial} reports raw MID-macro accuracy for Spatial-Omni on eligible FOA inputs. Audio-only accuracy is 31.3\% for Location and 50.5\% for Motion, above their respective 2.6\% and 25\% uniform-choice references. Radial accuracy is 50.3\%, near its 50\% reference, limiting competence-based interpretations of its conflict diagnostics. 

\begin{figure*}[t]
\centering
\includegraphics[width=\textwidth]{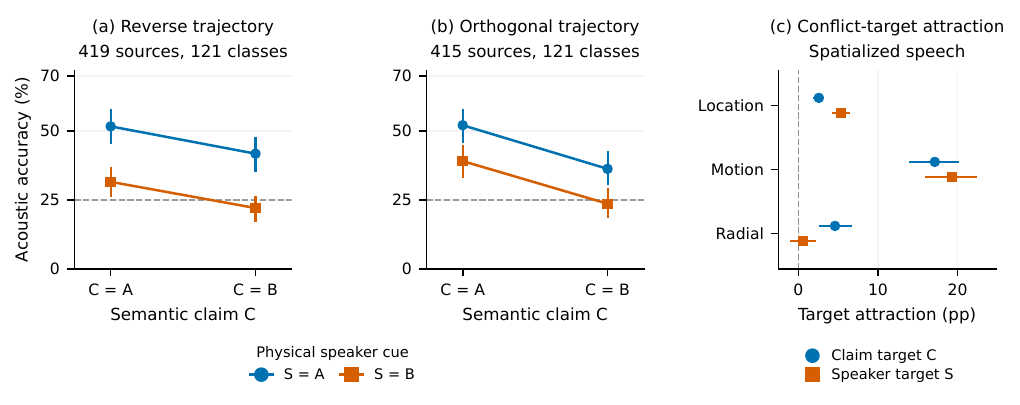}
\caption{Claim and speaker cues in Spatial-Omni's FOA motion task. (a,b) Matched acoustic accuracy when claim $C$ and speech trajectory $S$ match ground truth $A$ or an alternative $B$. (c) Attraction to conflicting $C$ and $S$ relative to audio-only controls. Error bars show 95\% class-bootstrap confidence intervals; dashed lines mark 25\% chance in (a,b) and zero attraction in (c).}
\label{fig:spatial_cues}
\end{figure*}

\paragraph{Semantic and physical interference.}
We analyze complete matched four-cell families with fixed environmental audio, crossing aligned or alternative semantic claims ($C$) and physical speech trajectories ($S$). Relative to $C=S=A$, Motion accuracy is 9.9 pp lower (95\% CI: [6.0, 13.7]) when only the semantic claim is reversed, and 20.1 pp lower ([15.2, 25.2]) when only the physical speech trajectory is reversed (Figure~\ref{fig:spatial_cues}a). For orthogonal trajectories, the corresponding differences are 15.8 pp ([11.7, 20.2]) and 13.0 pp ([8.6, 17.5]; Figure~\ref{fig:spatial_cues}b). Both cues therefore interfere with environmental trajectory judgments. Their relative point estimates vary across contrasts; these comparisons do not establish internal causal mechanisms.

\paragraph{Target attraction and joint conflict.}
With spatialized speech, Motion CTA-C and CTA-S are 17.1 and 19.3 pp, with both 95\% intervals above zero (Figure~\ref{fig:spatial_cues}c). Predictions thus shift toward both the spoken claim and the speech-source trajectory relative to matched audio-only controls. Radial CTA-S is only 0.6 pp (95\% CI: [-0.9, 2.1]), illustrating task-dependent attribution. Across all seven available factorial contrasts, interaction intervals include zero; the data do not establish nonadditive interference.

\subsection{Prompt Intervention and Conflict Robustness}
\label{sec:prompt_intervention}

\begin{figure*}[t]
\centering
\includegraphics[width=0.9\textwidth]{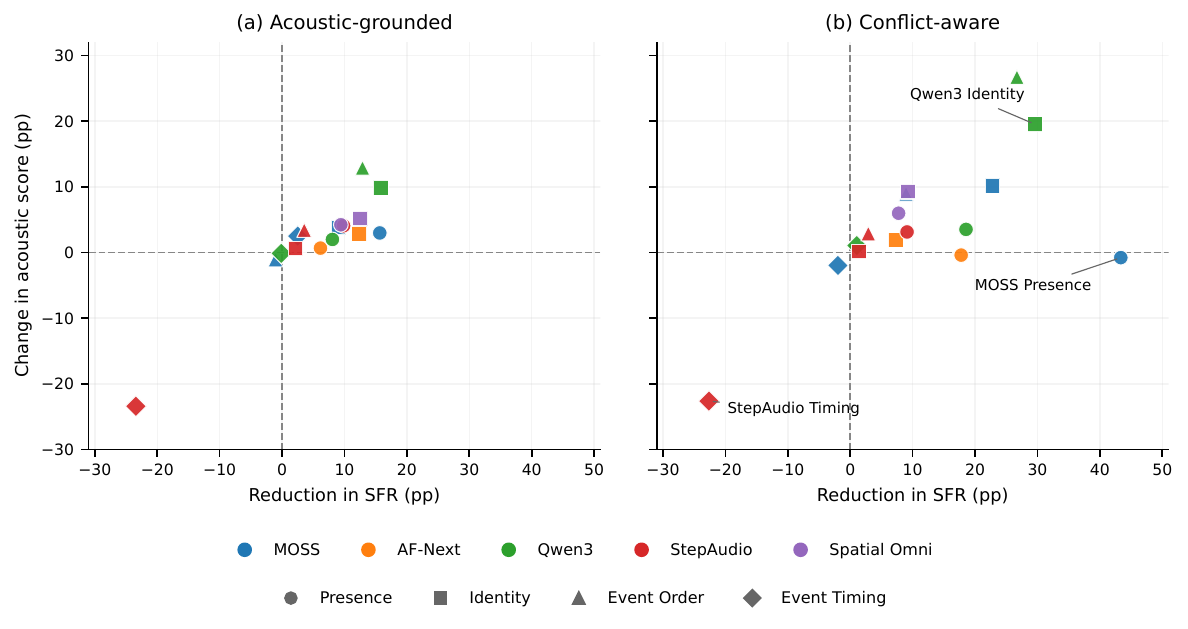}
\caption{Paired changes in acoustic performance and SFR under prompt interventions relative to neutral prompting. Rightward indicates reduced SFR; upward indicates improved acoustic performance. Presence uses recording-level BAcc. Model--task comparisons without eligible matched pairs are omitted.}
\label{fig:4_3_prompt_tradeoff}
\end{figure*}

\begin{table*}[t]
\centering
\small
\caption{Prompt-policy performance on representative conflict tasks (\%). Neu: neutral; AG: acoustic-grounded; CA: conflict-aware. Presence uses recording-level BAcc; other tasks use accuracy.}
\vspace{5pt}
\label{tab:4_3_prompts}
\begin{tabular}{llrrrrrr}
\toprule
 & & \multicolumn{3}{c}{Acoustic score (\%)} & \multicolumn{3}{c}{SFR (\%)} \\
\cmidrule(lr){3-5}\cmidrule(lr){6-8}
Model & Task & Neu & AG & CA & Neu & AG & CA \\
\midrule
MOSS-Audio-8B & Presence & 65.7 & 68.6 & 64.9 & 67.5 & 51.8 & 24.1 \\
\midrule
MOSS-Audio-8B & Identity & 29.1 & 32.9 & 39.2 & 60.9 & 51.8 & 38.1 \\
Audio Flamingo Next & Identity & 35.6 & 38.5 & 37.5 & 71.5 & 59.1 & 64.1 \\
Qwen3-Omni & Identity & 39.8 & 49.6 & 59.3 & 70.5 & 54.7 & 40.9 \\
Step-Audio-2-mini & Identity & 32.9 & 33.5 & 33.0 & 80.8 & 78.7 & 79.4 \\
Spatial-Omni (SO-30B) & Identity & 51.3 & 56.5 & 60.6 & 47.0 & 34.4 & 37.7 \\
\midrule
Qwen3-Omni & Temporal Order & 58.4 & 71.3 & 85.1 & 41.6 & 28.7 & 14.9 \\
\midrule
Step-Audio-2-mini & Temporal State & 78.9 & 55.5 & 56.3 & 21.1 & 44.5 & 43.7 \\
\bottomrule
\end{tabular}
\end{table*}

We compare neutral, acoustic-grounded (AG), and conflict-aware (CA) prompting with unchanged task questions and answer options (Table~\ref{tab:4_3_prompts}; Figure~\ref{fig:4_3_prompt_tradeoff}). We assess both acoustic correctness and speech-following behavior.

\paragraph{Selective recovery of acoustic grounding.} 

CA prompting improves Qwen3-Omni's Identity accuracy from 39.8\% to 59.3\%, while reducing SFR from 70.5\% to 40.9\%. On Temporal Order, accuracy rises from 58.4\% to 85.1\%, with 204 incorrect-to-correct repairs against only one correct-to-incorrect change in the matched comparison. MOSS-Audio and Spatial-Omni also gain 10.1 and 9.3 pp on Identity, respectively. However, Step-Audio-2-mini remains near 33\% Identity accuracy under all three policies. Prompting can therefore restore acoustically correct responses, but its effectiveness varies substantially across models.

\paragraph{Reduced speech following without improved grounding.} 
Under CA prompting, MOSS-Audio's Event Presence SFR falls from 67.5\% to 24.1\%, yet recording-level BAcc remains nearly unchanged (65.7\% to 64.9\%), with a paired 95\% confidence interval that includes zero. Class-specific recalls reveal the underlying response shift: Yes recall decreases from 69.9\% to 35.4\%, whereas No recall increases from 60.8\% to 94.5\%. Thus, the substantial reduction in speech following coincides with a pronounced preference for negative responses rather than improved balanced recognition of environmental events. This example shows that SFR suppression can arise from response redistribution without a corresponding gain in acoustic grounding.

\paragraph{Negative transfer under intervention.} 
Prompt interventions can also substantially worsen acoustic grounding. For Step-Audio-2-mini on Temporal State, accuracy decreases from 78.9\% under neutral prompting to 55.5\% with AG and 56.3\% with CA, while SFR increases from 21.1\% to 44.5\% and 43.7\%, respectively. Under AG prompting, 216 initially correct responses become incorrect, compared with only 38 repairs among 760 paired questions. This yields a -23.4 pp accuracy change, with a source-cluster bootstrap 95\% confidence interval of [-27.2, -19.5] pp. Notably, the task-specific instructions explicitly preserve speech as a temporal reference, yet performance still deteriorates. The results therefore reveal an intervention-induced failure pattern, although they do not establish its underlying mechanism. 

\paragraph{Limited changes in spatial grounding.}
For Motion with spatialized speech, neutral, AG, and CA prompting yield acoustic accuracies of 33.2\%, 32.9\%, and 33.6\%, respectively. CTA-C is 17.1, 13.9, and 15.0 pp, whereas CTA-S remains 19.3, 19.8, and 20.4 pp. The lower semantic-attraction point estimates therefore do not coincide with higher acoustic accuracy. These are descriptive policy-specific comparisons: the current spatial report does not provide paired cross-policy confidence intervals.

\paragraph{Scope of comparisons.}
Compatible--conflict gaps are condition-level aggregates, not paired instance effects. Spatial interference contrasts use matched families with fixed environmental audio, but speech waveforms vary across cells. Spatial results are limited to eligible Spatial-Omni FOA inputs; cross-policy differences remain descriptive because paired cross-policy confidence intervals are not reported. These analyses describe observable responses, not internal mechanisms. Synthesized speech and rendered spatial scenes may limit generalization to spontaneous speech and real-world environments.

\section{Conclusion}
\label{sec:conclusion}

We introduced SAC-Bench, a controlled benchmark for evaluating speech--acoustic conflict across semantic, temporal, and spatial grounding. Strong audio-only recognition does not guarantee robustness to conflicting speech, and Temporal Order failures occur even when speech follows the complete event sequence. Within the evaluated spatial setting, matched motion contrasts reveal accuracy losses associated with both spoken claims and physical speech trajectories. 

Prompt interventions yield heterogeneous outcomes: they can restore correct judgments, reduce speech following without improving accuracy, or degrade performance. These findings distinguish acoustic competence, susceptibility to competing speech cues, and intervention effectiveness as separate evaluation properties. SAC-Bench provides controlled conflicts and matched comparisons to assess these properties through both acoustic correctness and cue-aligned responses.

\clearpage
\section*{Acknowledgments}
We thank Jingze Zhang, Xiaoya Zhou and Yingpin Fang (Beijing Foreign Studies University), and Lee Sean Joe (National University of Singapore) for manually reviewing and refining ChatGPT-generated synonyms, including removing inappropriate candidates and adding missing alternatives.

\newpage
\bibliography{main}
\bibliographystyle{tmlr}

\clearpage
\appendix

\section{Dataset Construction Details}
\label{app:dataset_details}

Source recordings are curated independently within the original FSD50K development and evaluation splits. We first remove redundant ancestor labels using the AudioSet ontology, then discard annotations whose available PP/PNP ratings are not uniformly positive. Annotations without available ratings are retained. A recording is eligible only if exactly one event label remains after these filtering steps. All reported model evaluations use recordings originating from the exhaustively annotated FSD50K evaluation split.

\subsection{Duration Bins}

Environmental recordings are grouped into $[2,4)$, $[4,7)$, $[7,12)$, $[12,16.2)$, $[16.2,21.75)$, and $[21.75,31)$ seconds. Class eligibility and contrastive-label selection are computed independently within each FSD50K source split and duration bin.

\subsection{Contrastive Event Selection}
\label{app:event_selection}

Let $\mathbf e_i$ denote the CLAP text embedding of
event label $i$. Semantic distance is
\begin{equation}
d_{\mathrm{CLAP}}(i,j)
=
1-
\frac{\mathbf e_i^\top\mathbf e_j}
{\|\mathbf e_i\|_2\|\mathbf e_j\|_2}.
\end{equation}
For each target, $\mathcal C_i$ contains the
$k=\max(1,\lfloor0.2|\mathcal L_{s,b}|\rfloor)$
most distant eligible labels.

For AudioSet ontology node $v$, Seco intrinsic information
content is
\begin{equation}
IC(v)
=
1-
\frac{\log(|\mathrm{Desc}(v)|+1)}
{\log|\mathcal V|},
\end{equation}
where $\mathcal V$ is the ontology and
$\mathrm{Desc}(v)$ contains its descendants.
Lin similarity is
\begin{equation}
S_{\mathrm{ont}}(i,j)
=
\frac{2IC(\mathrm{LCA}(i,j))}
{IC(i)+IC(j)},
\end{equation}
where $\mathrm{LCA}(i,j)$ is the common ancestor with the
highest information content.

The far candidate is
\begin{equation}
j_i^{\mathrm{far}}
=
\arg\min_{j\in\mathcal C_i}S_{\mathrm{ont}}(i,j).
\end{equation}
Let $\mathcal R_i$ contain candidates having an
ancestor--descendant relation with $i$. The near candidate is
\begin{equation}
j_i^{\mathrm{near}}
=
\arg\max_{j\in\mathcal C_i\setminus\mathcal R_i}
S_{\mathrm{ont}}(i,j).
\end{equation}
If $\mathcal C_i\setminus\mathcal R_i$ is empty, the
maximization uses $\mathcal C_i$. Temporal-order construction instead samples one label uniformly from $\mathcal C_i$.

For semantic question construction, four additional distractor labels are selected independently of the near and far acoustic alternatives. We exclude the target label, the previously selected near and far labels, their duplicate display names, and candidates having an ancestor--descendant relationship with the target. The remaining eligible labels are sorted by ascending CLAP distance, with ties resolved by AudioSet class ID. Four distinct alternatives are then selected at the 12.5th, 37.5th, 62.5th, and 87.5th percentiles of this ordered candidate list, using deterministic nearest-position selection. These alternatives are used to construct question options rather than additional acoustic conflict conditions.

\subsection{Speech Synthesis}
\label{app:speech_synthesis}

Speech is synthesized using F5-TTS with two speaker identities, A and B, defined by separate reference waveforms and reference transcripts. Speaker-specific speed factors of 1.0 and 0.81672 are applied during synthesis. Each generated speaker segment is normalized to a target RMS of 0.055, with peak attenuation applied when necessary to prevent clipping. Multi-turn dialogue segments are concatenated in their original order and saved as 16-bit PCM audio at 44.1 kHz.

Table~\ref{tab:construction_configuration} records
configuration values needed to reproduce curation and
speech synthesis.

\begin{table}[t]
\centering
\small
\begin{tabular}{@{}p{0.59\columnwidth}p{0.33\columnwidth}@{}}
\toprule
\textbf{Configuration} & \textbf{Value} \\
\midrule
CLAP checkpoint  & 630k-audioset-best.pt \\
CLAP label-text template & Corrected human-readable label \\
Temporal-order sampling seed & 42 \\
GPT-4o model version & gpt-4o-2024-11-20 \\
F5-TTS checkpoint & F5TTS\_Base \\
M-AILABS reference recording & elliot\_miller \\
LJSpeech reference recording & Linda Johnson \\
Speaker-specific speed factors & Speaker A: 1.0; Speaker B: 0.81672 \\
Segment RMS target & $0.055$ \\
TTS inference seed & 4242 \\
TTS NFE steps & 32 \\
TTS CFG strength & 2.5 \\
TTS deterministic mode & enabled \\
TTS output sample rate & 44.1 kHz \\
\bottomrule
\end{tabular}
\caption{Dataset-generation configuration.}
\label{tab:construction_configuration}
\end{table}

\subsection{Spatial Coordinates and Conflict Conditions}
\label{app:conflict_details}

\paragraph{Spherical discretization.}
We use a listener-centered coordinate system with $+X$ pointing forward, $+Y$ leftward, and $+Z$ upward. Azimuth $\theta\in[-180^\circ,180^\circ)$ is measured from $+X$ toward $+Y$, and elevation $\phi\in[-90^\circ,90^\circ]$ is measured from the horizontal plane.

We construct 38 spherical Voronoi zones with the topology of the octahedral Goldberg polyhedron $GP_4(3,0)$~\citep{renka1997algorithm}. The generating sites are normalized permutations and sign variations of $(3,0,0)$, $(2,1,0)$, and $(1,1,1)$, yielding 6, 24, and 8 sites, respectively. Their spherical Voronoi dual contains six quadrilateral and 32 hexagonal regions.

For a unit direction $\mathbf u$, zone membership is
determined by
\begin{equation}
Z(\mathbf u)=
\arg\max_{i\in\{1,\ldots,38\}}
\mathbf u^\top\mathbf c_i,
\end{equation}
where $\mathbf c_i$ denotes the unit vector of site $i$. Source directions are sampled uniformly in solid angle within their assigned Voronoi cells using rejection sampling. The regions are not assumed to have equal area.

\paragraph{Static location.}
Let $A$, $C$, and $S$ denote the acoustic event direction, spoken claim direction, and physical speaker direction, respectively. For each acoustic anchor $A$, we construct five semantic conditions according to their angular separation:
\begin{equation}
\Delta(A,C)=
\cos^{-1}(\mathbf u_A^\top\mathbf u_C).
\end{equation}
These comprise Same ($\Delta=0^\circ$), Near
($0^\circ<\Delta\leq60^\circ$), Medium
($60^\circ<\Delta\leq120^\circ$), Far
($120^\circ<\Delta<180^\circ$), and Antipodal
($\Delta=180^\circ$). The four non-matching claims
occupy distinct zones.

For spatialized speech, five three-source relations
are defined:
\begin{align}
\mathrm{L1}:&\quad A=C=S,\\
\mathrm{L2}:&\quad A=C\neq S,\\
\mathrm{L3}:&\quad A=S\neq C,\\
\mathrm{L4}:&\quad C=S\neq A,\\
\mathrm{L5}:&\quad A\neq C,\ A\neq S,\ C\neq S.
\end{align}
The construction contains one L1, one L2, and four instances each of L3--L5. Together with five non-spatial speech conditions ($S$ undefined), this produces 19 conditions per source recording.

For L2, the speaker direction is selected from the four conflicting semantic directions, permitting an L2--L4 comparison at fixed $A$ and $S$. L3 and L4 share $A$ and $C$ while varying $S$. L5 selects $S$ from a separate set of Near, Medium, Far, and Antipodal candidates, excluding the zones occupied by $A$ and $C$. Candidate selection balances angular-distance strata across source families.

\paragraph{Spherical motion.}
We select two non-neighboring zones whose sampled directions are separated by $60^\circ$--$120^\circ$. The acoustic source follows the shorter great-circle arc between its endpoints. We construct three trajectory variants: the acoustic ground truth, its reversal, and a midpoint-orthogonal trajectory. The latter is obtained by rotating the entire arc by $\pm90^\circ$ about the radial axis through its midpoint. The trajectories have orthogonal tangents at the shared midpoint, rather than a constant $90^\circ$ angular separation.

The three variants define the possible semantic claims $C$. Physical speech motion $S$ independently follows one of the same three trajectories or remains non-spatial, producing $3\times4=12$ conditions.

\paragraph{Radial motion.}
For each fixed direction, the acoustic source either approaches the listener from 2.0 to 0.5\,m or recedes from 0.5 to 2.0\,m. Each acoustic direction is paired with both approaching and receding semantic claims. Speech is either non-spatial, moving in the same radial direction as the acoustic event, or moving in the opposite direction. This yields $2\times2\times3=12$ conditions.

\paragraph{Source sampling and matching.}
Within each FSD50K split and event class, we select five distinct source recordings across eligible duration bins using deterministic sampling. Selected WAVs must decode completely, contain finite non-silent audio, and match their declared duration bins. Recordings are not duplicated to fill a class quota.

Each selected recording is reused across its eligible matched spatial conditions. Clips shorter than 4 seconds contribute only to Static Location; longer clips additionally contribute to Spherical and Radial Motion. This preserves the acoustic material across controlled variations in $C$ and $S$.

Table~\ref{tab:benchmark_conditions} summarizes the
compatible and conflicting conditions.

\begin{table}[t]
\centering
\small
\begin{tabular}{@{}lll@{}}
\toprule
\textbf{Task} & \textbf{Compatible} & \textbf{Conflict} \\
\midrule
Presence (+) & target & none / near / far \\
Presence ($-$) & none / near / far & target \\
Identity (+) & target & none / near / far \\
Identity ($-$) & none / near / far & target \\
Temporal State & matched & reversed \\
Temporal order & matched & reversed \\
Location & $C=A$ & $C\neq A$ \\
Spherical motion & $C=A$ & $C\neq A$ \\
Radial motion & $C=A$ & $C\neq A$ \\
\bottomrule
\end{tabular}
\caption{Semantic compatibility conditions. In spatial tasks, semantic compatibility requires the spoken claim $C$ to match the acoustic ground truth $A$, independent of speaker geometry $S$.}
\label{tab:benchmark_conditions}
\end{table}

\section{Audio Rendering Details}
\label{app:audio_rendering}

\subsection{Semantic Rendering}
\label{app:semantic_rendering}

Speech waveform $s$ and environmental waveform $e$ are
temporally center-aligned. Their relative level is
\begin{equation}
\rho
=
20\log_{10}
\frac{\operatorname{RMS}(s)}
{\operatorname{RMS}(e)}.
\end{equation}
Speech level is fixed and event gain is chosen to attain the requested $\rho$. If peak protection is required, common attenuation is applied to the complete mixture. Speech-only and audio-only semantic controls reference the corresponding source waveforms without additional mixture rendering.

\subsection{Temporal Rendering}
\label{app:temporal_rendering}

For Temporal State, the environmental event is placed before or after the complete speech waveform with gap $\delta_s=0.2$ seconds.
For Temporal Order, two environmental waveforms are concatenated with an inter-event gap $\delta_e=0.2$ seconds, and speech follows the complete sequence after an additional gap $\delta_s=0.2$ seconds. Source recordings are assigned using a matched policy: for each event-class pair within the same source split and duration bin, the same recording for each event class is reused across compatible, conflicting, and audio-only conditions. Compatible and conflicting conditions use the same spoken claim but reverse the acoustic event order. Audio-only controls preserve the corresponding event sequence and inter-event gap while removing speech.

Level calibration uses pooled environmental RMS,
\begin{equation}
r_e
=
\sqrt{
\frac{\sum_{k=1}^{K}\sum_n e_k[n]^2}
{\sum_{k=1}^{K}N_k}
},
\end{equation}
where $N_k$ is the length of event $k$.
Inserted silence is excluded.
Speech gain is selected relative to $r_e$, and clipping
protection attenuates speech and events jointly.

\subsection{Spatial Rendering}
\label{app:spatial_rendering}

\paragraph{Source trajectories.}
Environmental events and spatialized speech are rendered using independently specified geometries. Static sources maintain a fixed direction at 2.0\,m. Spherical motion follows a shortest great-circle arc at constant radius, while radial motion maintains a fixed direction and varies distance linearly between 0.5 and 2.0\,m.

Moving sources remain stationary for 0.3\,s at each endpoint and move at constant angular or radial velocity during the intervening interval. Trajectories are sampled at 10-ms intervals.

\paragraph{Binaural rendering.}
We use SADIE II D2 KEMAR HRIRs in SOFA format \citep{armstrong2018perceptual}. Static sources undergo direction-specific convolution. Moving sources use blockwise time-varying convolution, with inverse-distance-weighted HRIR interpolation ($k=4$, exponent $2$) and a 256-sample block size. SOFA delay information is preserved to maintain interaural timing cues.

\paragraph{FOA encoding.}
We use first-order Ambisonics with AmbiX ACN/SN3D conventions~\citep{nachbar2011ambix}. For a mono source $x(t)$ at unit direction $\mathbf u(t)=[u_x,u_y,u_z]^\top$, the encoded signal is
\begin{equation}
[W,Y,Z,X]
=
g(r(t))x(t)[1,u_y,u_z,u_x],
\end{equation}
where $g(r)$ is the distance attenuation. Both environmental events and spatialized speech follow this encoding.

\paragraph{Distance and Doppler.}
Free-field distance attenuation is modeled as
\begin{equation}
g(r)=\frac{r_{\mathrm{ref}}}{r},
\end{equation}
with $r_{\mathrm{ref}}=1$\,m. Radial motion additionally uses retarded-time resampling:
\begin{equation}
t_r=t_e+\frac{r(t_e)-r(0)}{c},
\end{equation}
where $c=343$\,m/s is the speed of sound. This produces Doppler shifts consistent with the time-varying radial distance. Constant-radius spherical motion has no radial Doppler shift.

\paragraph{Speech spatialization.}
We distinguish two speech geometry modes. In the non-spatial mode, speech is added equally to both binaural channels or to the FOA $W$ channel only. No physical speaker direction is assigned.

In the spatialized mode, speech is rendered as an independent acoustic source with physical geometry $S$, using the same binaural or FOA rendering pipeline as the environmental event. The speaker's physical geometry is specified independently of the spoken semantic claim $C$. Both speech and event waveforms are temporally center-aligned before mixing.

\paragraph{Level calibration and controls.}
Speech level is calibrated after spatial rendering to a 0-dB speech-to-event ratio (SER). Active RMS is measured over non-overlapping 50-ms frames within the speech--event temporal overlap in which both sources exceed their respective activity thresholds of $-40$\,dB relative to peak frame RMS. A single gain is applied to the speech component:
\begin{equation}
g_s=
10^{\rho/20}
\frac{\operatorname{RMS}_{e}}
     {\operatorname{RMS}_{s}},
\qquad \rho=0\,\mathrm{dB}.
\end{equation}
Binaural calibration uses joint-ear energy, whereas FOA calibration uses the $W$ component.

Both formats use anechoic rendering without added synthetic noise. The original recording background, if present, is retained. For each distinct spatially rendered environmental event, an audio-only control is retained and reused across matching speech conditions.

\subsection{Rendering Parameters}
\label{app:rendering_parameters}

Table~\ref{tab:rendering_parameters} summarizes the
rendering configuration.

\begin{table}[t]
\centering
\small
\begin{tabular}{@{}p{0.59\columnwidth}p{0.33\columnwidth}@{}}
\toprule
\textbf{Parameter} & \textbf{Value} \\
\midrule
Speech--event gap & 0.2 s \\
Inter-event gap & 0.2 s \\
Speech position for Temporal order & after sequence \\
Semantic SER settings & $-10,-5,0,5,10$ dB \\
Acoustic environment & free-field \\
Spatial formats & binaural, FOA \\
Outer/static radius $R$ & 2.0 m \\
Near radius $a$ & 0.5m \\
Trajectory sampling interval & 10ms \\
Endpoint hold & 0.3 s \\
Reference distance & 1.0 m \\
Speed of sound & 343 m/s \\
Doppler & enabled \\
HRTF set & SADIE II D2 KEMAR \\
Moving-source interpolation & IDW, $k=4$ \\
IDW exponent & 2 \\
Convolution block size & 256 samples \\
Active-RMS frame & 50 ms \\
Activity threshold & $-40$ dB \\
Spatial speech modes & non-spatial, spatialized \\
Spatial additive noise & none \\
Spatial SER levels & 0 dB only \\
\bottomrule
\end{tabular}
\caption{Audio rendering parameters used in dataset generation.}
\label{tab:rendering_parameters}
\end{table}

\section{Extended Dataset Statistics}
\label{app:dataset_statistics}

The benchmark draws on 16,340 environmental recordings (45.9 hours) and 42,379 synthesized speech waveforms (93.7 hours), corresponding to 40,536 distinct speech texts. Its 616,031 unique final audio clips total 2,069.9 hours. Clips are deduplicated by normalized final audio path, not waveform-content hashing. Each path and its duration are counted once in the global total; all referenced final files passed the existence check.

\begin{table}[t]
\centering
\small
\caption{Unique audio clips and path-deduplicated duration by task.
Task rows overlap; the total is their union, not their sum.}
\label{tab:dataset_statistics_extended}
\begin{tabular}{lrr}
\toprule
Task & Unique clips & Hours \\
\midrule
Presence & 242,555 & 781.8 \\
Identity & 242,550 & 786.6 \\
Temporal State & 4,404 & 24.2 \\
Temporal Order & 6,562 & 54.9 \\
Static Location & 64,546 & 194.6 \\
Spherical Motion & 32,258 & 142.1 \\
Radial Motion & 34,694 & 121.7 \\
\midrule
All tasks (union) & 616,031 & 2,069.9 \\
\bottomrule
\end{tabular}
\end{table}

Of the unique clips, 558,333 contain speech and environmental audio, 28,580 are speech-only, and 29,118 are audio-only. Development contains 393,010 clips (1,326.1 hours), and evaluation contains 223,021 clips (743.8 hours). Environmental source paths, pair identifiers, synthesized speech identifiers, and final audio paths are disjoint across splits; 1,415 exact speech texts occur in both. Audio-clip counts are distinct from question and inference-request counts.

\section{Evaluation Implementation Details}

\subsection{Question Construction}
\label{app:question_construction}

\paragraph{Presence}
For recordings with environmental audio, queries include the acoustic event label and its registered alternatives. For recordings without an environmental event, the speech/target label anchors the alternative registry. Conflicting conditions additionally include the spoken event label when available. Repeated query labels are merged. A query receives acoustic answer Yes only when it identifies the actual event; all queries receive No in no-event recordings.

Speech-implied Yes/No references are assigned only to queries about the spoken target label when its claim defines the corresponding polarity. Queries about unrelated alternatives do not automatically receive a speech reference.

\paragraph{Identity}
Each question contains five distinct choices, with distractors drawn from the registered contrastive alternatives described in Appendix~\ref{app:event_selection}. The acoustic event, or ``No environmental sound is present'' for no-event audio, is always included. Positive conflicting claims additionally require the speech-asserted event. Remaining slots are filled from the registered alternatives, excluding duplicates of required choices. Alternative selection and final ordering are deterministic. Negative identity claims do not generally specify a unique speech-implied identity answer.

\paragraph{Temporal and radial motion}
Temporal State chooses between event-before-speech and event-after-speech. Temporal order chooses between the two possible orders of the specified events. Radial motion chooses approaching or receding.

\paragraph{Spatial questions.}
Static Location uses all 38 registered zones as distinct choices, with option keys A--Z and AA--AL. Spherical Motion asks for the complete directed trajectory using four distinct alternatives: the actual trajectory, its reversal, and its two midpoint-orthogonal rotations by $+90^\circ$ and $-90^\circ$. The bank includes both rotations even though each source family renders only one as a semantic alternative. Radial Motion uses approaching and receding. Each primary question has one correct key, with uniform-choice references of $1/38$, $1/4$, and $1/2$, respectively.

\paragraph{Spatial diagnostics and ordering.}
Neutral-prompt diagnostics query the spoken claim $C$, physical speech state $S$ when defined, and their relation to the environmental target $A$. Claim and physical-state questions use the corresponding primary bank. Relation questions distinguish L1--L5 for Location and Motion, or R1--R4 for Radial. Motion endpoints are separate 38-way diagnostics. All diagnostics are excluded from primary acoustic scores. The spatial build has no validated speech-only controls; mixed-input diagnostics do not measure independent single-source competence.

Spatial banks and permutations are fixed within each source family, independently of $C$, $S$, representation, and prompt policy. Mixed primary questions use all three policies; diagnostics and audio-only controls use neutral prompting. Nonspatial query and choice ordering follows the original deterministic seed-42 procedure.

\subsection{Question Templates}

The neutral policy adds no intervention instruction.
The model-facing questions are:

\begin{enumerate}
\item Presence:
``Is the sound of \{query label\} present in the recording?''
\item Identity:
``Which of the following best identifies the environmental
sound in the recording?''
\item Temporal State:
``When does the environmental sound occur relative to
the spoken segment?''
\item Temporal order:
``In what order do the two environmental sounds occur?''
\item Location:
``Which direction does the non-speech environmental sound come from,
relative to the listener?''
\item Spherical Motion:
``Which directed trajectory best describes the movement of the
non-speech environmental sound?''
\item Radial Motion:
``Is the non-speech environmental sound approaching or receding
from the listener?''
\end{enumerate}

Presence appends ``Answer only with Yes or No.''
Multiple-choice questions display the ordered options and
append ``Answer only with the corresponding letter.''

Spatial multiple-choice questions display their ordered choices and
append ``Answer only with the corresponding option key.''
Motion start and end questions are retained only as diagnostics.

\subsection{Intervention Instructions}
We report neutral prompting as the baseline and evaluate acoustic-grounded and conflict-aware prompting as two intervention conditions.

\paragraph{Presence and Identity}
\texttt{acoustic\_grounded}:
``Base your answer on the actual environmental sounds in the
recording. Do not use statements made by the speakers as
evidence about whether an environmental sound is present
or what that sound is.''

\texttt{conflict\_aware}:
``The spoken content may contain incorrect claims about the
environmental sounds in the recording. Treat such claims as
untrusted and independently determine the answer from the
actual environmental sounds.''

\paragraph{Temporal State}
\texttt{acoustic\_grounded}:
``Determine the answer from the actual timing in the
recording. Use the spoken segment only as a temporal
reference, not as evidence about when the environmental
sound occurs.''

\texttt{conflict\_aware}:
``The spoken content may incorrectly describe the timing of
the environmental sound. Use the spoken segment only as a
temporal reference and independently determine the event
timing from the recording.''

\paragraph{Temporal order}
\texttt{acoustic\_grounded}:
``Determine the order from the environmental sounds you
actually hear. Do not use statements made by the speakers as evidence about the event order.''

\texttt{conflict\_aware}:
``The spoken content may incorrectly describe the order of
the environmental sounds. Treat such claims as untrusted
and independently determine the actual order from the
recording.''

\paragraph{Spatial localization}
\texttt{acoustic\_grounded}:
``Determine the source location from the spatial acoustic
cues in the recording. Do not use statements made by the
speakers as evidence about the source location.''

\texttt{conflict\_aware}:
``The spoken content may incorrectly describe the location of
the environmental sound source. Treat such claims as untrusted
and independently determine the source location from the
spatial acoustic cues.''

\paragraph{Spherical motion}
\texttt{acoustic\_grounded}:
``Determine the source trajectory from the spatial acoustic
cues in the recording. Do not use statements made by the
speakers as evidence about the source trajectory.''

\texttt{conflict\_aware}:
``The spoken content may incorrectly describe the trajectory
of the environmental sound source. Treat such claims as
untrusted and independently determine the source trajectory
from the spatial acoustic cues.''

\paragraph{Radial motion}
\texttt{acoustic\_grounded}:
``Determine whether the sound source is approaching or
receding from the spatial acoustic cues in the recording.
Do not use statements made by the speakers as evidence
about the source motion.''

\texttt{conflict\_aware}:
``The spoken content may incorrectly describe whether the
sound source is approaching or receding. Treat such claims
as untrusted and independently determine the motion from
the spatial acoustic cues.''

\subsection{Inference Configuration}
\label{app:inference_configuration}

MOSS-Audio and Qwen3-Omni use vLLM with their model-specific audio interfaces and chat templates. Step-Audio-2-mini uses a vLLM-hosted chat-completion endpoint. Audio Flamingo Next uses Transformers with \texttt{MusicFlamingoForConditionalGeneration} in bfloat16. Spatial-Omni uses its official PyTorch implementation, Qwen3-Omni base model, adaptation checkpoint, and spatial audio encoder.

MOSS-Audio averages input channels before mono inference. Audio Flamingo Next reads float32 audio, averages channels, and uses polyphase resampling to 16 kHz. Step-Audio converts input to 16-kHz mono, 16-bit PCM WAV. Qwen3-Omni loads audio through \texttt{qwen\_omni\_utils.process\_mm\_info}; its processor behavior is tied to the recorded package version.

Spatial-Omni's FOA path requires four channels, preserves their order during polyphase resampling to 16 kHz, and uses a maximum 20-second input window. Binaural requests and overlength recordings are explicitly skipped rather than truncated or substituted with mono audio.

Table~\ref{tab:inference_configuration} records the
effective configuration of the reported runs.

\begin{table*}[t]
\centering
\caption{Model checkpoints and inference configurations for the reported experiments.}
\small
\begin{tabular}{@{}p{0.34\textwidth}
                    p{0.16\textwidth}
                    p{0.11\textwidth}
                    p{0.10\textwidth}
                    p{0.17\textwidth}@{}}
\toprule
\textbf{Checkpoint} & \textbf{Backend version} & \textbf{Precision} & 
\textbf{Max. tokens} & \textbf{Revision} \\
\midrule
MOSS-Audio-8B-Instruct &
vLLM: 0.31.0 & BF16 & 256 & 66326e6 \\  
Spatial-Omni-SO30B &
PyTorch: 2.9.0 & BF16 & 256 & ba33a4c \\
audio-flamingo-next-hf &
Transformers: 5.19.0 & BF16 & 256 & 5634886 \\
Qwen3-Omni-30B-A3B-Instruct &
vLLM: 0.9.3 & BF16 & 256 & 26291f7 \\
Step-Audio-2-mini &
vLLM: 0.1.dev1+g9ecabf22f & BF16 & 256 & e36fdd5 \\
\bottomrule
\end{tabular}

\label{tab:inference_configuration}
\end{table*}

Requests are assigned to deterministic logical shards by resolved audio path, keeping questions and policies for the same recording together. Stored outputs retain request identifiers, task metadata, reference sets, raw responses, and execution signatures.

\subsection{Answer Parsing and Outcome Accounting}
\label{app:answer_scoring}

Nonspatial results use the relaxed parser. Spatial results use a separate final-line choice parser supporting both one- and two-letter option keys. Closed reasoning blocks and special-token markup are removed; an unclosed reasoning block is unparseable. Accepted spatial forms include a standalone displayed key, supported answer wrappers, and a key accompanied by its matching displayed option text. Text-only, fuzzy, or synonym-based spatial answer matching is not used.

Completed invalid or unparseable responses count as incorrect. Inference errors, missing predictions, and explicit skips are excluded from primary accuracy and reported separately. The reported Spatial-Omni run contains 327,084 requests across primary questions, diagnostics, and prompt policies: 132,585 completed responses (132,420 valid and 165 invalid), 163,542 unsupported-format skips, and 30,957 overlength skips, with no missing predictions or inference errors. These are request counts, not unique-audio or independent-source counts.

\subsection{Presence Metrics}
\label{app:presence_metrics}

For recording $r$, let $\mathcal Q_{r,y}$ contain completed queries
with acoustic reference $y\in\{\mathrm{Yes},\mathrm{No}\}$.
Define
\begin{equation}
R_{r,y}=
\frac{1}{|\mathcal Q_{r,y}|}
\sum_{i\in\mathcal Q_{r,y}}
\mathbf{1}[\hat y_i=y].
\end{equation}
The primary score is
\begin{equation}
\mathrm{BAcc}_{\mathrm{wav}}=
\frac{1}{|\mathcal R_{\pm}|}
\sum_{r\in\mathcal R_{\pm}}
\frac{R_{r,\mathrm{Yes}}+R_{r,\mathrm{No}}}{2},
\end{equation}
where $\mathcal R_{\pm}$ contains recordings with completed queries of both reference classes. All queries are used, without negative subsampling; recordings receive equal weight.

Invalid completed responses remain incorrect. Recordings losing a completed reference class through inference failures are excluded from BAcc and counted separately. For analysis conditions with only one acoustic reference class by construction, the primary score is recording-averaged recall for that class, not BAcc. Question-pooled recalls and BAcc are supplementary diagnostics.

\subsection{Spatial Scoring and Matched Diagnostics}
\label{app:chance_metrics}

\paragraph{Primary aggregation and uncertainty.}
Let $a_i$ denote correctness on a completed environmental-target question.
The primary score is
\begin{equation}
\mathrm{Acc}_{\mathrm{sp}}=
\frac{1}{|\mathcal M|}\sum_{m\in\mathcal M}
\frac{1}{|\mathcal R_m|}\sum_{r\in\mathcal R_m}
\frac{1}{|\mathcal Q_{mr}|}\sum_{i\in\mathcal Q_{mr}}a_i,
\end{equation}
where $\mathcal M$ contains contributing event-class MIDs, $\mathcal R_m$ their original environmental recordings, and $\mathcal Q_{mr}$ the completed primary questions for a recording. Primary scores are raw accuracy. The supplementary adjustment is $(\mathrm{Acc}_{\mathrm{sp}}-c)/(1-c)$, with fixed $c=1/38$, $1/4$, or $1/2$ for Location, Motion, or Radial.

Spatial 95\% intervals resample MIDs with replacement while retaining their within-class recording groups. They are conditional on the selected recordings. Nonspatial intervals instead use environmental-source clusters; Temporal Order uses connected components of shared sources. Both procedures use 1,000 bootstrap replicates with base seed 42. Paired contrast intervals are computed from paired differences.

\paragraph{Matched factorial contrasts.}
Hold environmental state $A$ fixed within a family and select an
alternative $B$.
Let $p_{ij}$ denote accuracy with $C=A$ for $i=0$ and $C=B$ for $i=1$,
and $S=A$ for $j=0$ and $S=B$ for $j=1$.
The aligned-other-cue interference contrasts and interaction are
\begin{align}
\Delta_C &= p_{00}-p_{10},\\
\Delta_S &= p_{00}-p_{01},\\
I_{CS} &= p_{11}-p_{10}-p_{01}+p_{00}.
\end{align}
Only families with four completed responses, the same source, and identical choices contribute. Differences are formed within families before source/MID aggregation. Location contributes only distance contrasts with a complete four-cell structure; not every source supplies all distance contrasts. These contrasts are descriptive because speech waveforms can differ. An interaction interval containing zero does not establish equivalence.

\paragraph{Conflict-target attraction.}
For target role $T\in\{C,S\}$, let $T_i$ denote its spatial value for question $i$, restricting eligibility to $T_i\ne A_i$. Let $\hat v_i^{\mathrm{mix}}$ and $\hat v_i^{A}$ denote the spatial values selected on a mixture and its unique matched neutral audio-only control. Both responses must be completed and share environmental source, spatial state, representation, and ordered choices. Define
\begin{equation}
d_{i,T}=
\mathbf{1}[\hat v_i^{\mathrm{mix}}=T_i]-
\mathbf{1}[\hat v_i^{A}=T_i].
\end{equation}
CTA-T is the source/MID macro-average of $d_{i,T}$. The control need not be correct. Invalid completed responses contribute zero to target indicators. CTA-C and CTA-S can use different eligible sets and share a target when $C=S\ne A$; CTA alone does not separate the two mechanisms.

\paragraph{Conditional acoustic retention.}
Retention is mixed-input primary accuracy restricted to matched audio-only controls answered correctly, with the same source/MID aggregation. Correct-control selection does not establish competence when baseline performance is near chance. Spatial prompt-policy scores remain descriptive because the current report supplies no paired cross-policy intervals.

\end{document}